\documentclass[%
 reprint,
showpacs,preprintnumbers,
 amsmath,amssymb,
 aps,
pra,
]{revtex4-1}

\usepackage{graphicx}% Include figure files
\usepackage{dcolumn}% Align table columns on decimal point
\usepackage{bm}% bold math
\usepackage{color}
\usepackage{upgreek}
\begin{document}

\preprint{APS/123-QED}

\title{Scalable Focused Ion Beam Creation of Nearly Lifetime-Limited\\ Single Quantum Emitters in Diamond Nanostructures}
\author{Tim Schr\"oder$^{\dagger}$}
 \altaffiliation{Now at Niels Bohr Institute, University of Copenhagen, DK.}

\author{Matthew E. Trusheim}%
\altaffiliation[These authors contributed equally ]{}
\author{Michael Walsh}
\altaffiliation[These authors contributed equally ]{}
\author{Luozhou Li}
\author{Jiabao Zheng}
\altaffiliation[Also at ]{Columbia University, New York, NY, 10027, USA.}
\author{Marco Schukraft}
\author{Dirk Englund}
\email{schroder@nbi.ku.dk; englund@mit.edu}
\affiliation{%
 Department of Electrical Engineering and Computer Science, Massachusetts Institute of Technology, Cambridge,
MA 02139, USA}
\author{Jose L. Pacheco}
\author{Ryan M. Camacho}
\author{Edward S. Bielejec}
\affiliation{
 Sandia National Laboratories, Albuquerque, NM 87185, USA}
\author{Alp Sipahigil}
\author{Ruffin E. Evans}
\author{Denis D. Sukachev}
\altaffiliation[Also at ]{Russian Quantum Center and P.N. Lebedev Physical Institute, Moscow, 143025, Russia.}
\author{Christian T. Nguyen}
\author{Mikhail D. Lukin}
\affiliation{Department of Physics, Harvard University, 17 Oxford St., Cambridge, MA 02138, USA}

\date{\today}

\begin{abstract}
The controlled creation of defect center---nanocavity systems is one of the outstanding challenges for efficiently interfacing spin quantum memories with photons for photon-based entanglement operations in a quantum network. Here, we demonstrate direct, maskless creation of atom-like single silicon-vacancy (SiV) centers in diamond nanostructures via focused ion beam implantation with $\sim 32$~nm lateral precision and $< 50$~nm positioning accuracy relative to a nanocavity. Moreover, we determine the Si+ ion to SiV center conversion yield to $\sim 2.5\%$ and observe a 10-fold conversion yield increase by additional electron irradiation. We extract inhomogeneously broadened ensemble emission linewidths of $\sim$51~GHz, and close to lifetime-limited single-emitter transition linewidths down to $126 \pm13$~MHz corresponding to $\sim 1.4$-times the natural linewidth. This demonstration of deterministic creation of optically coherent solid-state single quantum systems is an important step towards development of scalable quantum optical devices.

\end{abstract}

\pacs{Valid PACS appear here}
\maketitle

%\tableofcontents

A central goal in semiconductor quantum optics is to devise efficient interfaces between photons and atom-like quantum emitters for applications including quantum memories, single photon sources, and nonlinearities at the level of single quanta. Many approaches have been investigated for positioning emitters relative to the mode-maximum of nanophotonic devices with the necessary sub-wavelength-scale precision, including fabrication of nanostructures around pre-localized or site-controlled semiconductor quantum dots (QDs) \cite{badolato_deterministic_2005,sapienza_nanoscale_2015,schneider_microcavity_2012,birindelli_single_2014,surrente_polarization_2015} or diamond defect centers \cite{riedrich-moller_nanoimplantation_2015}; or implantation of ions for defect center creation in nanostructures concomitant with the nanofabrication \cite{schroder_targeted_2014,schukraft_invited_2016}. However, these approaches have not allowed post-fabrication creation of quantum emitters with nearly indistinguishable emission in nanophotonic structures already fabricated and evaluated; such an approach would greatly simplify the design and fabrication process and can improve the quantum emitter---nanostructure fabrication yield. Recently, a method for incorporating nitrogen vacancy (NV) centers into diamond cavities was demonstrated by implanting nitrogen ions through a pierced atomic force microscope (AFM) \cite{riedrich-moller_nanoimplantation_2015}; however, new methods are required to scale this approach to the sample- or wafer-scale and to produce  indistinguishable photon emission with spectral linewidth close to the inverse excited-state lifetime. 

Here, we introduce a new method for positioning emitters relative to the mode-maximum of nanophotonic devices: focused-ion beam implantation of Si atoms into diamond photonic structures. This post-fabrication approach to quantum emitter generation achieves nanometer-scale positioning accuracy and creates SiV centers with optical transition linewidths comparable to the best naturally incorporated SiV reported \cite{rogers_multiple_2014}. The approach allows Si implantation into $\sim2\cdot10^4$ sites/s, which allows creation of millions of emitters across a wafer-scale sample. We also show that additional post-implantation electron irradiation and annealing creates an order of magnitude enhancement in Si to SiV conversion yield. By repeated cycles of Si implantation and optical characterization, this approach promises nanostructures with precisely one SiV emitter per desired location. Finally, we demonstrate and evaluate the site-targeted creation of SiVs in pre-fabricated diamond photonic crystal nanocavities. The ability to implant quantum emitters with high spatial resolution and yield opens the door to the reliable fabrication of efficient light-matter interfaces based on semiconductor defects coupled to nanophotonic devices. 

The SiV belongs to a group of color centers in diamond that have emerged as promising single photon emitters and spin-based quantum memories. Among the many diamond-based fluorescent defects that have been investigated \cite{aharonovich_diamond_2014}, the silicon-vacancy (SiV) center \cite{wang_single_2006,neu_single_2011,neu_photophysics_2012,neu_low_2013} is exceptional in generating nearly lifetime-limited photons with a high Debye-Waller factor of 0.79 \cite{collins_annealing_1994} and low spectral diffusion due to a vanishing permanent electric dipole moment in an unstrained lattice \cite{sipahigil_indistinguishable_2014,hepp_electronic_2014}. These favorable optical properties have notably enabled two-photon quantum interference between distant SiV centers \cite{sipahigil_indistinguishable_2014,rogers_multiple_2014} and from two SiV centers coupled to the same waveguide \cite{sipahigil_integrated_2016}. In addition, the SiV has electronic and nuclear spin degrees of freedom that could enable long-lived, optically-accessible quantum memories \cite{muller_optical_2014,pingault_all-optical_2014,rogers_all-optical_2014}.

\begin{figure}
	\includegraphics[width = 7 cm]{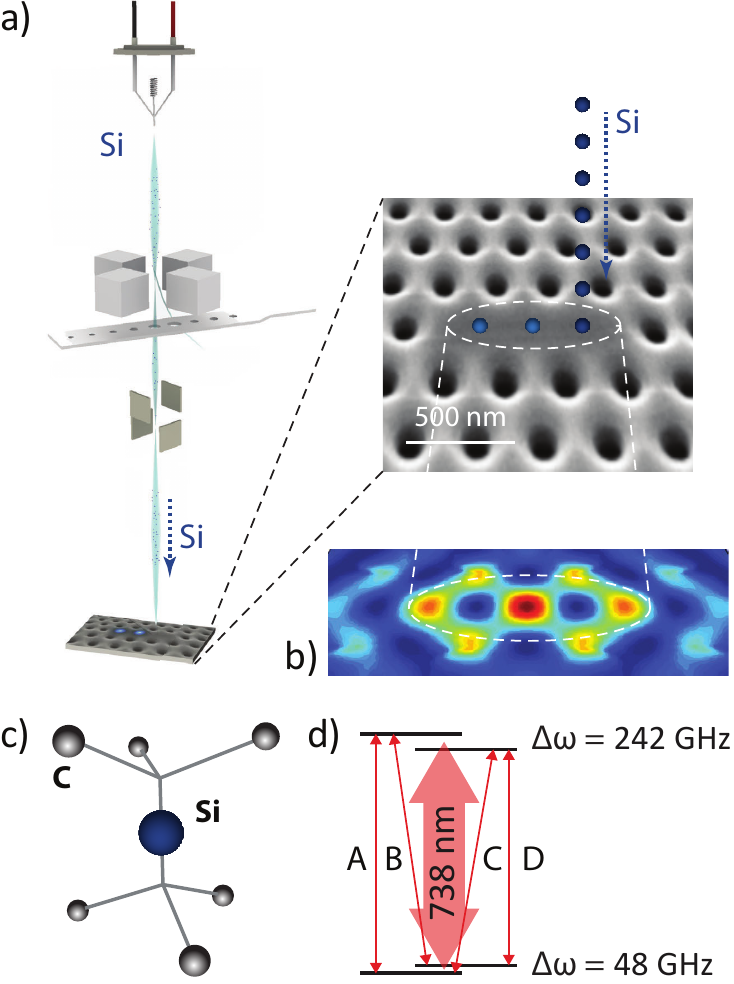}
	\caption{a) Illustration of targeted ion implantation. Si ions are precisely positioned into diamond nanostructures via a focused ion beam (FIB). The right side shows a scanning electron micrograph of a L3 photonic crystal cavity patterned into a diamond thin film. b) Intensity distribution of the fundamental L3 cavity mode with three Si target positions: the three mode maxima as indicated by the dashed lines. c) Atomic structure of a silicon vacancy defect center (SiV) in diamond. Si represents an interstitial Si atom between a split vacancy along the $<$111$>$ lattice orientation and C the diamond lattice carbon atoms.  d) Simplified energy level diagram of the negatively charged SiV indicating the four main transitions \cite{hepp_electronic_2014}. }
	\label{fig:illustration}
\end{figure}

Unlike quantum emitters such as molecules or quantum dots, diamond defect centers can be created through ion implantation and subsequent annealing \cite{meijer_generation_2005,rabeau_implantation_2006}, enabling  
direct control of the center depth via the ion energy.  
Lateral control has been demonstrated through the use of nanofabricated implantation masks \cite{naydenov_engineering_2009,toyli_chip-scale_2010,pezzagna_nanoscale_2010,spinicelli_engineered_2011,pezzagna_creation_2011,bayn_generation_2015}, which have been employed for color center creation relative to optical structures through atomic force microscope (AFM) mask alignment \cite{riedrich-moller_nanoimplantation_2015}, and combined implantation/nanostructure masking \cite{schroder_targeted_2014,schukraft_invited_2016}. Implantation through a pierced AFM tip \cite{riedrich-moller_nanoimplantation_2015} does not require modification of the fabrication process but is particularly time-consuming, requires special AFM tips,  and can lead to reduced positioning precision by collisions with mask walls. 
As an alternative, FIB implantation of ions, for example nitrogen \cite{lesik_maskless_2013} and silicon \cite{tamura_array_2014}, can greatly simplify the implantation process by eliminating the requirement of a nanofabricated mask. Similar to a scanning electron microscope, an ion beam can be precisely raster-scanned, enabling lateral positioning accuracy at the nanometer scale and `direct writing' into tens of thousands of structures with high throughput. \\

\begin{figure}
	\includegraphics[width = 8.5 cm]{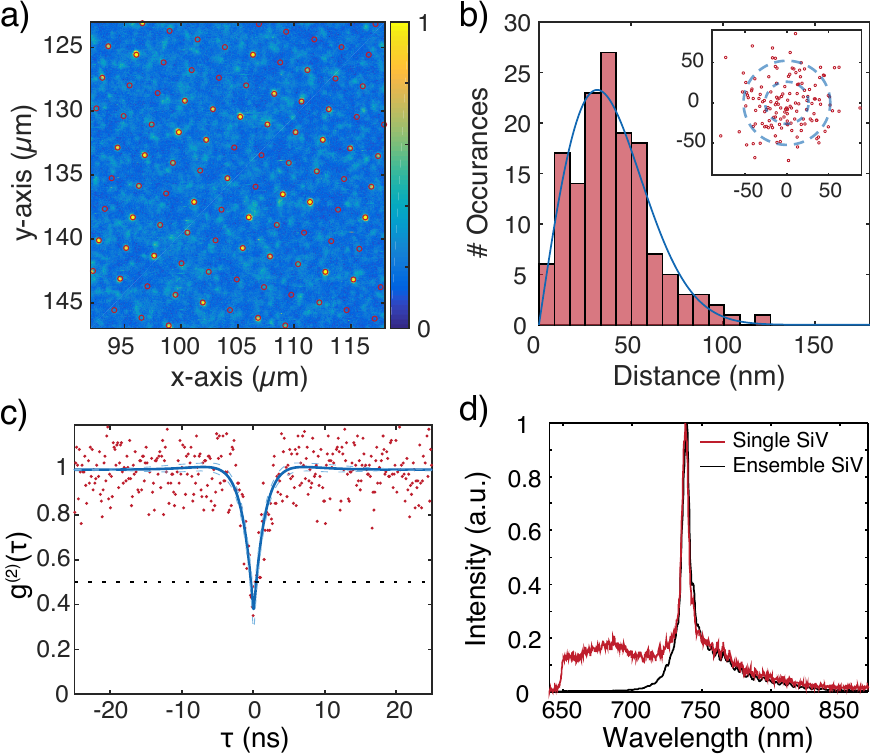}
	\caption{a) Confocal scan of SiV center array. Sites are separated by 2.14~$\upmu$m. Overlaid are regular grid points from an aberration-corrected reference lattice.  b) Analysis of implantation precision. We determine the 2-d position uncertainty of the created SiV be $40 \pm 20$nm. Red curve: fit to Rayleigh distribution. Inset: Scatter plot of created SiV sites relative the their grid points with one and two $\sigma$ guides to the eye, where the radius $\sigma=26$~nm corresponds to the expected implantation standard deviation resulting from the combination of beam width and implant straggle. c) Normalized second-order autocorrelation function of a single SiV with g$^{(2)}(0) = 0.38\pm0.09$. Blue points indicate data (without background subtraction), and the red line is a fit to the function $1-A\cdot exp(-|\tau/t_1|)+B\cdot exp(-|\tau/t_2|)$. The black dashed line indicates g$^{(2)}(\tau) = 0.5$ while the red dashed lines indicate the 95$\%$ confidence interval on the fit. d) Ensemble (blue) and single-emitter (red) SiV room-temperature fluorescence spectra. The characteristic zero-phonon line at 737~nm is prominent. }
	\label{fig:spectrum}
\end{figure}

\noindent
\textbf{Results}\\
\noindent
As outlined in Figure 1, the fabrication approach introduced here relies on Si implantation in a custom built 100 kV FIB nanoImplanter (A\&D FIB100nI) system (Appendix~\ref{sec:app1}) and subsequent high-temperature annealing to create SiV centers. The nanoImplanter uses field emission to create a tightly focused ion beam down to a minimum spot size of $< 10$~nm from a variety of liquid metal alloy ion sources (Appendix~\ref{sec:app1}). We applied commercially available high-purity chemical vapor deposition (CVD) diamond substrates (Element6). For the experiments described here, we used an Si beam with a typical spot size of $< 40$~nm into commercially available high-purity chemical vapor deposition (CVD) diamond substrates (Element6). After implantation, we performed high-temperature annealing and surface preparation steps to convert implanted Si ions to SiVs (Appendix~\ref{sec:app2}).

We characterized the resulting SiV arrays at room temperature through confocal fluorescence microscopy in a home-built setup (Appendix~\ref{sec:app3}). Fig.~\ref{fig:spectrum}a shows a scan of a square array of SiV implantation sites with lattice spacing of 2.14~$\upmu$m across a 30$\times$30~$\upmu$m$^2$ write field, created via a single point exposure from the Si beam.
Room-temperature spectral measurements in a dense region containing many centers (Fig \ref{fig:spectrum}d, blue curve) showed an inhomogeneous linewidth of approximately 5~nm centered around 738.3~nm, characteristic of the SiV center. We subsequently identified  single SiVs through second-order correlation measurements. For instance, Fig.~\ref{fig:spectrum}c shows photon antibunching for a SiV  with an observed count rate of 30~kcts/s collected via an oil immersion (numerical aperture of 1.3) objective into a single-mode fiber under 20~mW of 532~nm pump power. The red line in Fig.~\ref{fig:spectrum}d shows the single-emitter fluorescence spectrum at room temperature, which is very similar in shape and linewidth to the inhomogeneous spectrum. At room temperature, these lines are determined by phonon processes and not limited by inhomogeneity between different SiV centers \cite{jahnke_electronphonon_2015}. 

To determine the spatial precision of creating SiV with our method, we analyze their distribution relative to the implantation lattice grid. Fig.~\ref{fig:spectrum}b shows the distance of each imaged SiV implantation from the ideal lattice site, resulting in a $\chi$ distribution with a mean separation in R of $\sigma = 40 \pm 20$~nm and underlying lateral (x,y) distributions with zero mean and standard deviations of 32~nm. These measured values agree well with the expected precision of 26 nm calculated by the addition in quadrature of the uncertainties arising from the nominal 40 nm FHWM beam size and 19 nm lateral implantation straggle.

\begin{figure}
	\includegraphics[width = 8cm]{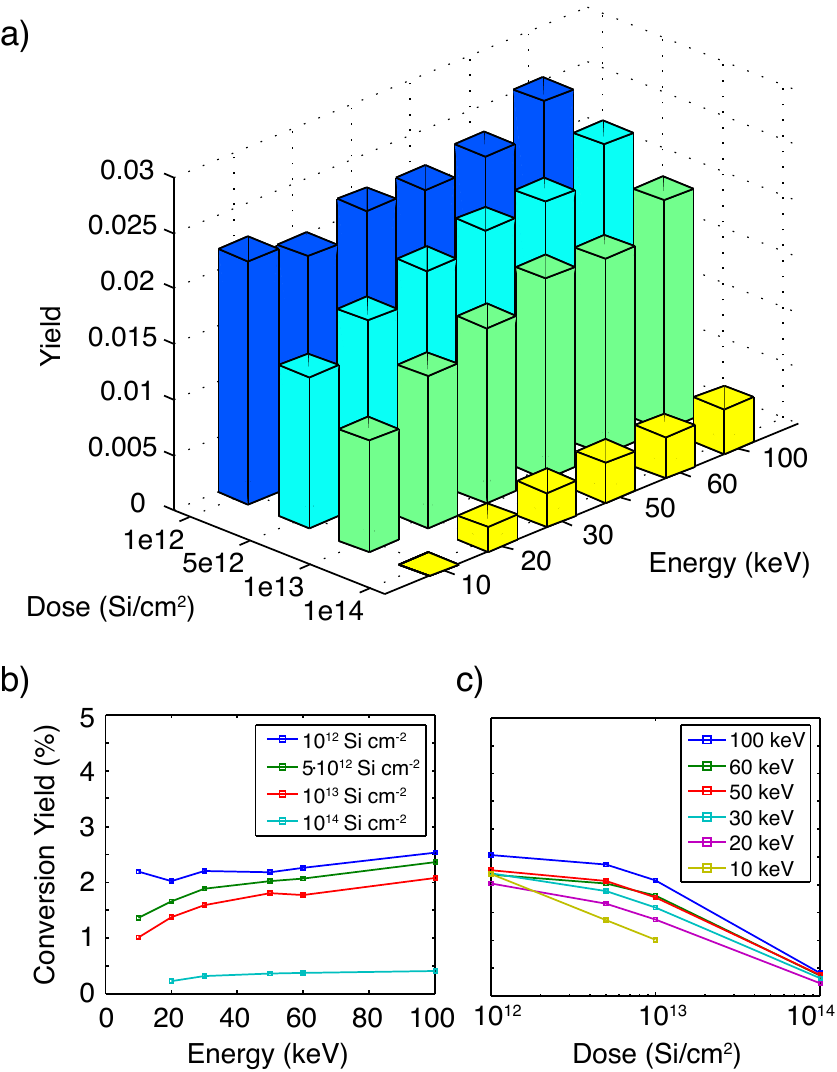}
	\caption{a) Si to SiV conversion yield for varying implantation ion energies and doses. The conversion yield was determined by calibrating array intensities (Fig.~\ref{fig:spectrum}) with the determined averaged single SiV photon count rate.
    Si conversion yield as function of b) energy and c) dose. The lines are guides to the eye.}
    \label{fig:yield}
\end{figure}

To determine the conversion yield of implanted Si ions to SiV centers, we swept the implantation dose logarithmically from $10^{12}$ to $10^{14}$ Si cm$^{-2}$, and the implantation energy linearly from 10 to 100~keV. The dose and energy determine the number and depth of vacancies created during the implantation process, with increased energy resulting in more vacancies at increased depth. This affects the probability that a Si defect captures a diffusing vacancy and converts to SiV during annealing which is a proposed mechanism for SiV formation \cite{dhaenens-johansson_optical_2011}. To estimate the yield, we measured the fluorescence intensity across a 10$\times$10~$\upmu$m$^2$ region of constant implantation dose and energy, and normalized to the average single-emitter intensity and implanted ion number. Fig.~\ref{fig:yield}a summarizes the yield measurements. Yield increases as a function of energy (Fig.~\ref{fig:yield}b), which is expected for a vacancy-limited SiV creation process, up to 2.5\% for the highest-energy 100~keV ions with a dose of $\sim$$10^{12}~\textrm{cm}^{-2}$. These measurements indicated a decreasing yield as a function of dose (Fig.~\ref{fig:yield}c). We attribute this to an accumulation of charged defects in the diamond lattice that lead to ionization, similarly to what was observed in NV centers \cite{schwartz_situ_2011}. Alternatively, reduced yield could result from lattice damage that accumulates in the form of multi-vacancy defects as the diamond lattice approaches the graphitization threshold, a phenomenon that has been observed in similar experiments with NV centers \cite{olivero_characterization_2006}. 

\begin{figure}
	\includegraphics[width = 8.5 cm]{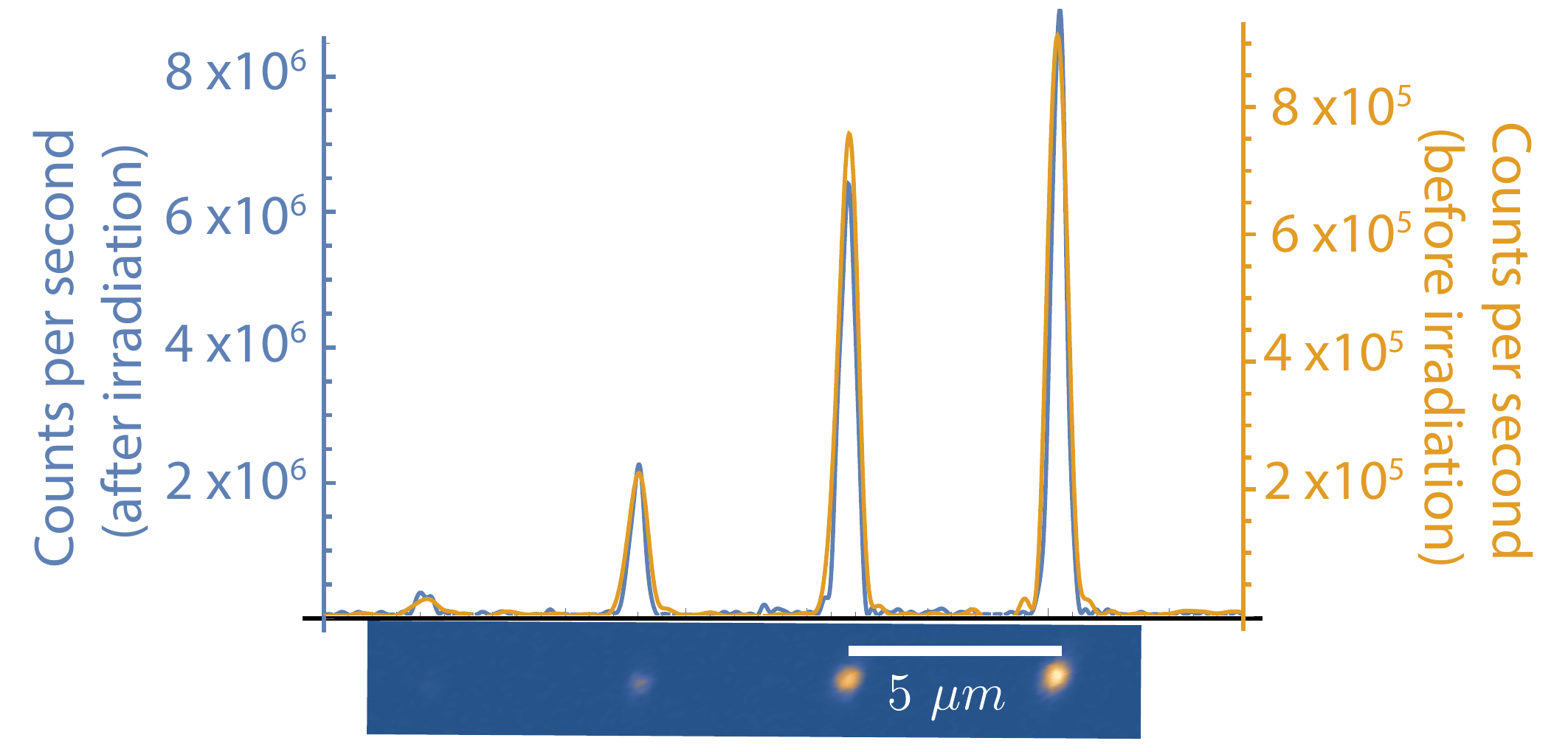}
	\caption{Electron co-implantation. After electron irradiation and subsequent annealing we observe a 10-fold increase in  fluorescence intensity at the implantation positions of Si ions (lower inset). The Si ion doses were 500, 2000, 5000, and 10000 ions per spot. The yellow line plot through the fluorescence maximum of the image indicates the intensity before electron irradiation, and the blue line after irradiation.
    }
	\label{fig:e-irrad}
\end{figure}

Irradiating diamond with high energy electrons can also improve the conversion yield of vacancy-related color centers \cite{martin_generation_1999,jarmola_longitudinal_2015}. Electron irradiation at high energies $>170$keV \cite{koike_displacement_1992} can displace carbon atoms and create additional vacancies, which allows for larger conversion efficiency of implanted ions into vacancy-related color centers. To verify these experiments with the silicon vacancy center, we first created a reference sample by implanting four spots with silicon ions in increasing doses of 500, 2000, 5000, and 10000 ions per spot into bulk diamond. After annealing this sample at 1200~$^{\circ}$C to activate SiVs \cite{davies_vacancy-related_1992}, a scanning confocal fluorescence image was taken by exciting these spots simultaneously with $\sim 10$~mW of both 520~nm (Thorlabs LP520-SF15) and 700~nm (Thorlabs LP705-SF15) laser light, and collecting light into a single mode fiber through a 10~nm bandpass filter (Semrock FF01-740/13) around 737~nm (Fig.~\ref{fig:e-irrad}, yellow line). After this reference measurement, we irradiated the sample with 1.5~MeV electrons with a total fluence of $\sim$$10^{17}~\textrm{cm}^{-2}$. After another annealing step, a second fluorescence image was taken with the same setup and it was verified by spectral measurements (Horiba iHR 550 with Synapse CCD) that indeed only the SiV typical peak at 737~nm was detected. In the second measurement, we observed increased fluorescence for all four spots by a factor of $\sim 10$ (Fig.~\ref{fig:e-irrad}, blue line), corresponding to a final conversion yield of $\sim 20\%$. This result is consistent with previous observations in Si-doped diamond samples \cite{dhaenens-johansson_optical_2011}, supporting our interpretation that the conversion efficiency of focused ion beam implantation is limited by the vacancy density in the diamond. 

\begin{figure}
	\includegraphics[width = 8.5cm]{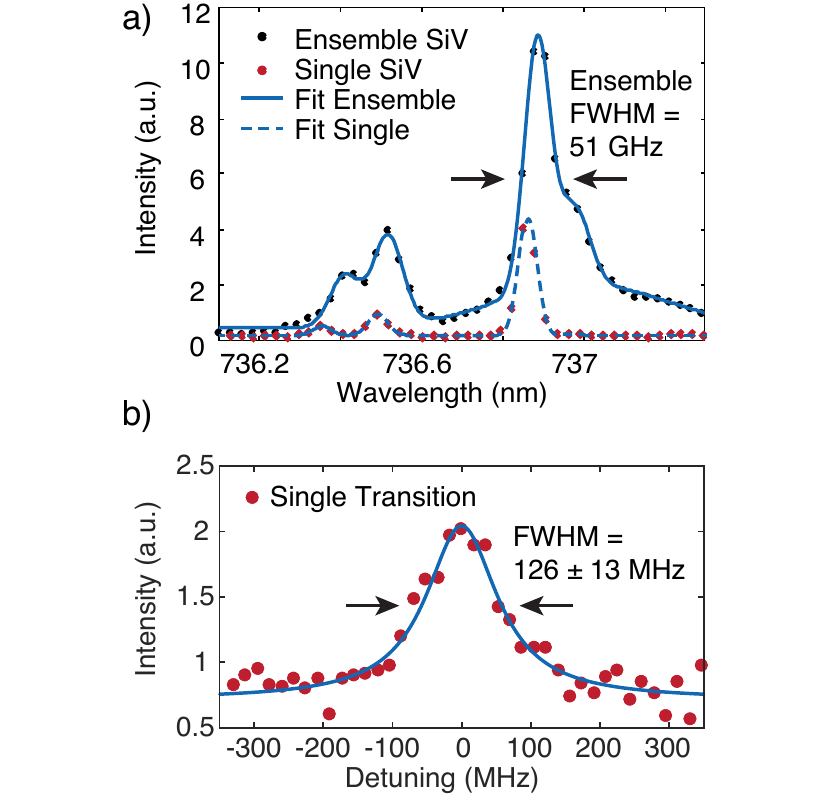}
	\caption{a) Cryogenic spectra ($<13~$K) of a single SiV (red circles) and an ensemble (black circles). The four SiV transitions (Fig.~\ref{fig:illustration}) as well as the phonon sideband are each fitted with a Gaussian function. The single SiV linewidths are spectrometer limited  ($\textrm{FWHM}=\sim 34$~GHz). 
    For the ensemble, we determine inhomogeneously broadened linewidths as low as $\sim$51~GHz (FWHM). The wavelength values are slightly blue-shifted due to an offset relative to an absolute wavelength reference by about 0.1~nm.
    b) Cryogenic ($4~$K) photoluminescence excitation measurement of the narrowest observed single SiV transition with a linewidth of $126\pm13$~MHz (FWHM) determined with a Lorentzian fit function. This linewidth of an implanted SiV is equal, within error, to the narrowest natural SiV linewidth measured to date.}
	\label{fig:cryo}
\end{figure}

We next describe the implanted SiV centers' low-temperature spectral properties. Photoluminescence spectral measurements were performed in a home-built confocal cryostat setup (Appendix~\ref{sec:app4}). The inhomogeneous distribution of SiV transitions is plotted in  Fig.~\ref{fig:cryo}a with a full width at half maximum (FWHM) of about 0.642~nm ($\sim51$~GHz). We then performed PLE measurements to determine the linewidths of individual SiVs below the spectrometer limit (Appendix~\ref{sec:app4}). We determined an average single-emitter transition linewidth of $200\pm15$~MHz from a sample of 10 SiV implanted at 100~keV with individually resolvable transitions. The narrowest observed transition, shown in Fig.~\ref{fig:cryo}b, had a linewidth of 126$\pm$13 MHz, which is within a factor of 1.4 of the lifetime limit $\gamma=(2\pi \cdot 1.7~\textrm{ns})^{-1}=94$~MHz for a typical fluorescence lifetime of 1.7~ns \cite{sipahigil_indistinguishable_2014}, and equivalent to the narrowest lines observed in natural SiVs to date \cite{rogers_multiple_2014,evans_narrow-linewidth_2016}. 

\begin{figure}
	\includegraphics[width = 8.5 cm]{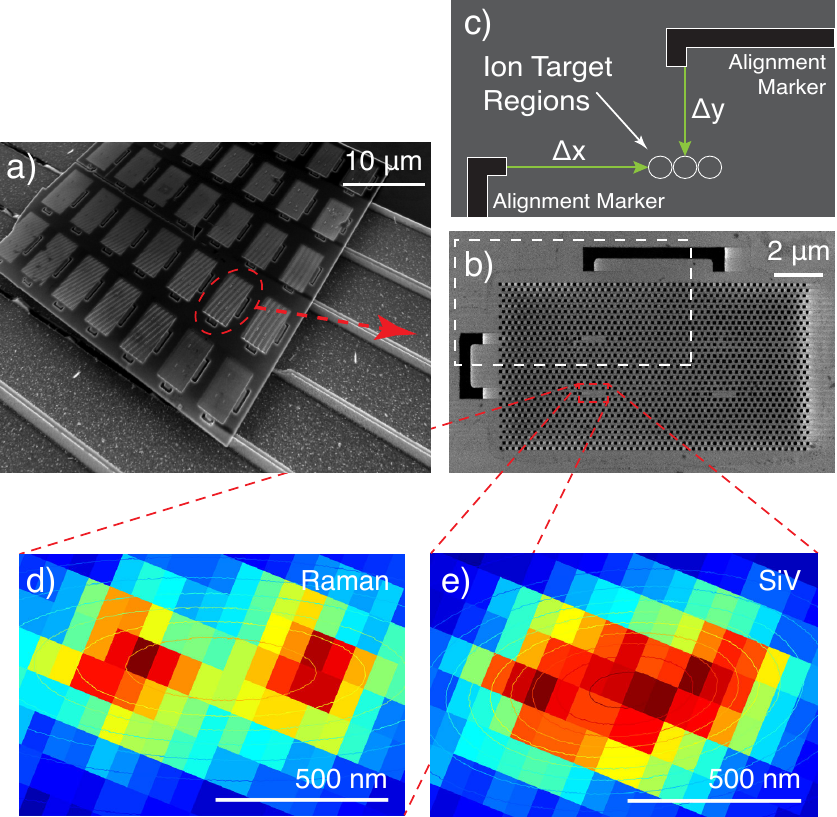}
	\caption{a) SEM of example PhC cavity sample. b) Close-up SEM of example PhC lattice with four cavities. The white dashed rectangle indicates the area illustrated in c). c) Illustration of targeting relative to alignment markers (black) with an ion spot size down to $<40$~nm. The white circles (not to scale for visibility) indicate the three L3 cavity mode maxima (Fig.~\ref{fig:illustration}). To determine the SiV positioning accuracy relative the the mode maxima, we performed spectrally resolving fluorescence scans. At each pixel in d,e) we recorded a spectrum including the Raman signal (d), the SiV fluorescence (e), and the cavity resonances (not displayed). d) Intensity x-y plot of the diamond Raman signal at 572.8~nm. e) Intensity x-y plot of SiV emission at 736.9~nm. By fitting a 2d Gaussian function to the intensity distribution, we determined the distance between the center of the cavity and the SiV fluorescence, the effective positioning accuracy, to $48(21)$~nm, with error estimation of one standard deviation.
 	}
	\label{fig:target}
\end{figure}

Finally, we demonstrated the targeted implantation and subsequent creation of SiV centers inside diamond nanostructures.  
We first fabricated 2D photonic crystal nanocavities into a $\sim$200~nm thick diamond membrane by oxygen reactive ion etching \cite{li_coherent_2015,li_nanofabrication_2015}. We then used the FIB system to target Si ions into the mode-maxima of the photonic crystal cavities. In the case of L3  cavities, we targeted the three mode-maxima individually (Fig.~\ref{fig:illustration}b). The Si ion beam was aligned to the cavity through secondary-electron imaging of pre-fabricated alignment markers on the sample (Fig.~\ref{fig:target}b, App.~\ref{sec:app1}). 
We targeted the cavity mode maxima with 160~keV ions for an average of 1.8~SiVs per cavity (App.~\ref{sec:app2}). After performing the processing steps described in Appendix~\ref{sec:app2} we observed about one SiV per cavity implantation spot with spectrometer-limited ($<34$~GHz) ZPL linewidths. To determine the position of a single SiV relative to the cavity, we performed a spectrally resolved photoluminescence confocal scan (Fig.~\ref{fig:target}d,e). This measurement allows comparison between the photonic crystal cavity location, determined by Raman scattering, and the SiV location, determined from the emitted fluorescence at the SiV ZPL. By fitting the measured emission patterns to 2-d Gaussians, we estimate a relative positioning accuracy of $ 48\pm21$~nm. This is close to the limit set by the combination of beam size and implantation straggle ($\sim$52~nm), with very low offset of 4~nm in the x-direction and a main offset of 48~nm in the y-direction. \\

\noindent
\textbf{Discussion}\\
\noindent
While we have demonstrated targeted creation of high-quality SiVs through FIB, there are several avenues for improvement. The stochastic creation yield, $\eta$, of defect centers, even with 10-fold yield increase by electron irradiation, prohibits the deterministic fabrication of single emitters which could be relevant for the high yield device fabrication \cite{mouradian_scalable_2015}. 
One solution is to implant a low dose of Si atom (to create $\ll$ 1 SiV on average) and optically verify if a SiV resulted after annealing. Due to the ability to select implantation sites individually, the FIB process allows for such repeated low-yield implantation steps conditionally halted on the creation of the desired emitter number. An alternative approach to create precisely one quantum emitter is to implant only one ion at a time, as was recently demonstrated \cite{abraham_fabrication_2016}, combined with electron irradiation or co-implantation of other ion species to create vacancies \cite{schwartz_situ_2011} to drive the SiV conversion yield to unity.

A remarkable property of the created SiVs is their narrow linewidth compared to the NV center, which has linewidths of several tens of gigahertz for a dose of $\sim$$10^{12}~\textrm{cm}^{-2}$ and a conversion yield of a few percent \cite{chu_coherent_2014}. The linewidths of the SiVs were measured in areas with on average 2.5 SiVs, distributed within only $\sim$55.4~nm (FWHM) diameter, corresponding to an implantation dose of $\sim$$10^{12}~\textrm{cm}^{-2}$, indicating that high densities of implanted SiVs are not detrimental for their optical properties. 

Although we found that FIB-implanted SiVs are similar in homogeneous transition linewidth to 'natural', as-grown centers, the inhomogeneous linewidth of $\sim$51~GHz is still slightly broader than the $\sim$15~GHz  demonstrated for a similar SiV creation method with annealing temperatures around 1200$^{\circ}$C \cite{evans_coherent_2015}. Potential causes include that (i) the higher temperature causes di-vacancies break down, or (ii) near-surface strain and defects in the diamond due to polishing, which can be reduced by etching the damaged layer before implantation \cite{chu_coherent_2014}.\\

In summary, we demonstrated SiV creation with high spatial accuracy by FIB implantation of Si atoms into bulk and nanostructured diamond. We show a SiV positioning accuracy relative to the mode maximum of a photonic crystal cavity of $ 48\pm21$~nm, which is sufficiently precise to allow implantation within the mode-field maximum of nanocavities or waveguides. We also demonstrate that the SiV creation yield can be increased after implantation by a factor of 10, important steps for reliable integration of quantum defects into on-chip photonic networks. The targeted implantation technique demonstrated here likely applies also to other quantum emitters and materials of interest, such as silicon carbide \cite{lienhard_bright_2016} or molybdenum disulfide; this would be particularly advantageous for materials for which traditional nanofabricated masking is challenging.

Remarkably, the ZPLs of SiVs created by our method have optical linewidths within a factor of 1.4 of the lifetime limit, making them as narrow as naturally occurring SiVs.  
Considering both this narrow linewidth and the narrow inhomogeneous distribution of implanted SiV of only $\sim$51~GHz, this fabrication method represents a significant step towards  the high-yield generation of thousands to millions of efficiently waveguide-coupled indistinguishable single photon sources. Such arrays of atom-like quantum emitters would be of great utility for a range of proposed quantum technologies, including quantum networks and modular quantum computing \cite{benjamin_prospects_2009,nemoto_photonic_2014}, linear optics quantum computing \cite{kok_linear_2007,li_resource_2015}, all-photonic quantum repeaters \cite{azuma_all-photonic_2015,pant_rate-distance_2016}, and photonic Boson sampling \cite{bryan_t._gard_introduction_2015}.

%-----------------------------------------

\section*{Appendix}

\subsection{\label{sec:app1}Silicon ion implantation}
Focused ion implantation was performed at the Ion Beam Laboratory at Sandia National Laboratories using the nanoImplanter (nI). The nI is a 100~kV focused ion beam (FIB) machine (A\&D FIB100nI) making use of a three-lens system designed for high mass resolution, using an ExB filter, and single ion implantation, using fast beam blanking.  The ExB mass-filter (M/ $\Delta$M of $\sim 61$) separates out different ionic species and charge states from liquid metal alloy ion sources (LMAIS), providing the capability for implantation of $\sim 1/3$ the periodic table over a range of energies from 10 to 200 keV.  For the Si implantation discussed here, we used an AuSbSi LMAIS with typical Si beam currents ranging from 0.4~pA to 1~pA. Fast beam blanking allows direct control over the number of implanted ions.  We determine the number of implanted ions by measuring the beam current and setting the pulse length to target a given number of ions per pulse. The nI is a direct write lithography platform that uses electrostatic draw deflectors, controlled by a Raith Elphy Plus pattern generator, to position the beam.  Single ion positioning is limited by the beam spot size on target.  With typical spot sizes ranging from 10-50 nm, we have measured the targeting accuracy to be $<35$ nm for 200keV~Si++ beam using a series of ion beam induced charge measurements.

For targeting into nanostructures, we align the ion beam relative to the sample by registering a secondary electron image of the alignment markers generated using the ion beam to scan the sample. Shift, rotation, and magnification corrections are calculated and applied in the pattern generator control package.  This allows for any location within the write field to be individually targeted for implantation. 

The lithography pattern is the original design file that was used to pattern the diamond thin film via electron beam lithography (EBL) and reactive ion etching. Errors resulting from inaccuracy during EBL were not taken into account. 

To create a single SiV per cavity with high probability, we implanted $\sim$20 Si ions per cavity mode maximum, yielding about 1.8~SiVs per cavity on average according to an extrapolated conversion efficiency of $\sim 3\%$ under Poisson statistics for 160~keV Si ions (Fig.~\ref{fig:yield}) that target the middle of membrane at 106~nm. 

\subsection{\label{sec:app2}SiV creation and sample preparation}
We annealed the sample at 1050$^{\circ}$C under high vacuum ($< 10^{-6}$~mbar at max temperature) for two hours to form SiV centers and eliminate other vacancy-related defects. Finally, we clean the sample surface through boiling tri-acid treatment (1:1:1 nitric:perchloric:sulfuric) and subsequent dry oxidation in a 30$\%$ oxygen atmosphere at 450$^{\circ}$C for four hours. 

\subsection{\label{sec:app3}Room Temperature Measurement Setup}
We employ a modified fluorescence microscope (Zeiss Axio Observer), customized to allow confocal illumination at 532 nm (Coherent Verdi) and single-mode fiber fluorescence collection. Collected fluorescence is spectrally filtered (Thorlabs FEL0650) and detected on avalanche photodiodes (APD) with single-photon resolution (Excelitas) or spectrally resolved on a grating spectrometer (Princeton Instruments, Acton SP2500i). 

\subsection{\label{sec:app4}Cryogenic Measurement Setup}
These measurements were performed at 18 K in a closed cycle helium cryostat (Janis). A home-built confocal microscope collects the fluorescence with a high numerical aperture (NA) objective (Olympus UMplanfl 100x 0.95NA) and directs the emission to either the input of a single-mode fiber connected to an APD or to a free-space spectrometer with a resolution of about 61~pm ($\sim 34$~GHz) at 737~nm (Princeton Instruments, IsoPlane SCT 320).

PLE measurements were performed using a modified helium flow probe-station (Desert Cryogenics model TTTP) with a 0.95 NA microscope objective (Nikon CFI LU Plan Apo Epi 100x) inside the vacuum chamber. Details of this setup are described in References\cite{evans_narrow-linewidth_2016}.

\subsection{\label{sec:app5a}Analysis of Spatial Positioning Precision}
To determine the spatial precision of the SiV implantation, we created and imaged a square array of SiV color centers following the procedures in appendicies A and D. We then fit each SiV site with a 2d Gaussian to determine the location of the SiV centers below the diffraction limit, and considered only SiV sites with fluorescence intensities consistent with single emitters. Using these locations, we fit a 2d grid allowing for affine transformation and find the distance between each SiV site and its nearest grid point. Finally, we bin the distances and fit to a central Chi distribution with two degrees of freedom (Rayleigh distribution), which describes the distribution of the distance $R = \sqrt[]{X^2+Y^2}$ where X and Y are independent zero-mean normal random variables with identical variance (Fig.~\ref{fig:spectrum}b). The reported separation is the mean of the fitted Chi distrbution corresponding to the mean separation in R (40~nm), and the error is the square root of the variance (20~nm). The mean separation in the X and Y directions is 0~nm with a standard deviation of 32~nm.

\subsection{\label{sec:app5}Analysis of Targeted Implantation Accuracy}
To determine the positioning accuracy of the cavity-targeted SiV creation, we performed a spectrally resolved photoluminescence confocal scan at room temperature. At each pixel of a 2-dimensional (2d) 532~nm laser scan we recorded a spectrum and determined the intensity for different spectral positions. For each wavelength, we then plotted its 2d-intensity map as in Fig.~\ref{fig:target}~d,e. This measurement allows comparison between the photonic crystal location, determined by Raman scattering of the 532~nm laser pump from the diamond (572.52~nm) which is present in the cavity region but not in the surrounding air holes, and the SiV location determined from the emitted fluorescence (at 736.98~nm). By fitting the measured emission patterns to 2d Gaussians, we estimate a relative positioning accuracy of $ 48(21)$~nm. The error is estimated from the 68\% fitting confidence interval which corresponds to one standard deviation.

\subsection{Acknowledgements}
The authors would like to thank Daniel L. Perry for his assistance in performing the implantation. This research was supported in part by the Army Research Laboratory Center for Distributed Quantum Information (CDQI), the CUA, the NSF, the AFOSR MURI, the Center for Integrated Quantum Materials (NSF grant DMR-1231319) and the DARPA QuINESS program. Ion implantation was performed at Sandia National Laboratories (SNL) with support from the Laboratory Directed Research and Development (LDRD Program and the Center for Integrated Nanotechnologies, an Office of Science (SC) facility operated for the U.S. DOE SC (contract DE-AC04-94AL85000) by the Sandia Corporation, a subsidiary of Lockheed Martin.

\bibliography{TimLibrarySiV5}

\end{document}